\documentclass[preprint, 12pt, superscriptaddress, amsmath, amssymb, showkeys]{revtex4-2}

\usepackage{bm}
\usepackage{booktabs}
\usepackage[dvipsnames]{xcolor}
\usepackage{graphicx}
\graphicspath{{./Figures}}
\usepackage[colorlinks]{hyperref}
\usepackage{layouts}

\usepackage{hyperref}

\renewcommand{\vec}{\bm}

\newcommand{\trans}{\ensuremath ^\mathsf{T}}
\newcommand{\fig}{Fig.~}
\newcommand{\eq}{Eq.~}
\newcommand{\tab}{Tab.~}
\newcommand{\sect}{Sec.~}
\newcommand{\etal}{\emph{et al.}}

\newcommand{\spacedim}{\ensuremath{d}}
\newcommand{\sysdim}{\ensuremath{D}}

\newcommand{\spacelen}{\ensuremath{L}}
\NewDocumentCommand{\dir}{o}{%
  \ensuremath{\hat{e}_{\IfValueTF{#1}{#1}{i}}}%
}

\newcommand{\measspace}{\ensuremath{k}}

\NewDocumentCommand{\meanof}{o}{%
  \ensuremath{\langle{\IfValueTF{#1}{#1}{}}\rangle}%
}
\NewDocumentCommand{\varof}{o}{%
  \ensuremath{\mathrm{var}(\IfValueTF{#1}{#1}{})}%
}

\newcommand{\nodes}{\ensuremath{N}}

\newcommand{\degree}{\ensuremath{\kappa}}
\newcommand{\spec}{\ensuremath{\rho}}
\newcommand{\inscale}{\ensuremath{\nu}}

\newcommand{\reg}{\ensuremath{\beta}}
\newcommand{\bin}{\ensuremath{b_{\mathrm{in}}}}
\newcommand{\bout}{\ensuremath{b_{\mathrm{out}}}}
\newcommand{\dt}{\ensuremath{\Delta t}}

\newcommand{\ttrans}{\ensuremath{t_{\mathrm{trans}}}}
\newcommand{\trainsteps}{\ensuremath{m_{\mathrm{train}}}}
\newcommand{\ttrain}{\ensuremath{t_{\mathrm{train}}}}

\newcommand{\wout}{\ensuremath{\boldsymbol{W}_{\mathrm{out}}}}
\newcommand{\wadj}{\ensuremath{\boldsymbol{W}_{\mathrm{adj}}}}
\newcommand{\win}{\ensuremath{\boldsymbol{W}_{\mathrm{in}}}}

\NewDocumentCommand{\res}{o}{%
  \ensuremath{\vec{s}_{\IfValueTF{#1}{#1}{m}}}%
}
\NewDocumentCommand{\inp}{o}{%
  \ensuremath{\vec{u}_{\IfValueTF{#1}{#1}{m}}}%
}
\NewDocumentCommand{\outp}{o}{%
  \ensuremath{\vec{y}_{\IfValueTF{#1}{#1}{m}}}%
}
\NewDocumentCommand{\iv}{o}{%
  \ensuremath{\vec{z}_{\IfValueTF{#1}{#1}{m}}}%
}
\NewDocumentCommand{\esv}{o}{%
  \ensuremath{\vec{x}_{\IfValueTF{#1}{#1}{m}}}%
}

\newcommand{\tvalid}{\ensuremath{t_{\mathrm{val}}}}

\newcommand{\ressamples}{\ensuremath{K}}

\newcommand{\esvmat}{\boldsymbol{X}}
\newcommand{\outmat}{\boldsymbol{Y}}

\newcommand{\indim}{{\ensuremath{D_{\mathrm{in}}}}}

\newcommand{\fper}{\ensuremath{f_{\mathrm{per}}}}

\newcommand{\parres}{\ensuremath{M}}
\newcommand{\parind}{\ensuremath{i}}
\NewDocumentCommand{\inppar}{o}{%
  \ensuremath{\inp[{\IfValueTF{#1}{#1}{m}}}]^{\,(\parind)}}
  
\NewDocumentCommand{\ivpar}{o}{%
  \ensuremath{\iv[{\IfValueTF{#1}{#1}{m}}}]^{\,(\parind)}}
\NewDocumentCommand{\outppar}{o o}{%
  \ensuremath{\outp[{\IfValueTF{#1}{#1}{m}}}]^{\,({\IfValueTF{#2}{#2}{\parind}})}}
\NewDocumentCommand{\corevar}{o o}{%
  \ensuremath{\inp[{\IfValueTF{#1}{#1}{m}}}]^{\,(\IfValueTF{#2}{#2}{i}, \mathrm{c})}}
\NewDocumentCommand{\neivar}{o}{%
  \ensuremath{\inp[{\IfValueTF{#1}{#1}{m}}}]^{\,(\parind, \mathrm{n})}}

\newcommand{\coredim}{{\ensuremath{D_{\mathrm{c}}}}}
\newcommand{\neidim}{{\ensuremath{D_{\mathrm{n}}}}}
\newcommand{\neilen}{{\ensuremath{l}}}

\newcommand{\neiind}{{J}}

\NewDocumentCommand{\winpar}{o}{%
  \ensuremath{\ensuremath{\win^{\,(\IfValueTF{#1}{#1}{\parind})}}}}
\NewDocumentCommand{\wadjpar}{o}{%
  \ensuremath{\ensuremath{\wadj^{\,(\IfValueTF{#1}{#1}{\parind})}}}}
\NewDocumentCommand{\woutpar}{o}{%
  \ensuremath{\ensuremath{\wout^{\,(\IfValueTF{#1}{#1}{\parind})}}}}
\NewDocumentCommand{\wincore}{o}{%
  \ensuremath{\ensuremath{\boldsymbol{W}_{\mathrm{c}}^{\,(\IfValueTF{#1}{#1}{\parind})}}}}
\NewDocumentCommand{\winleft}{o}{%
  \ensuremath{\ensuremath{\boldsymbol{W}_{\mathrm{l}}^{\,(\IfValueTF{#1}{#1}{\parind})}}}}
\NewDocumentCommand{\winright}{o}{%
  \ensuremath{\ensuremath{\boldsymbol{W}_{\mathrm{r}}^{\,(\IfValueTF{#1}{#1}{\parind})}}}}
\newcommand{\order}{\ensuremath{\boldsymbol{P}}}
\newcommand{\proc}{\ensuremath{\boldsymbol{\Pi}_{\mathrm{core}}}}

\newcommand{\dimred}{\ensuremath{\eta}}
\newcommand{\prodr}{\ensuremath{\boldsymbol{\Pi}_{\dimred}}}
\newcommand{\lintrafo}{\ensuremath{\mathcal{L}}}

\NewDocumentCommand{\ident}{o}{%
  \ensuremath{\boldsymbol{I}_{\IfValueTF{#1}{#1}{n}}}%
}

\newcommand{\real}{\ensuremath{\mathbb{R}}}

\begin{document}
\title{ Improving the prediction of spatio-temporal chaos by combining parallel reservoir computing with dimensionality reduction }

\author{Luk Fleddermann}
\author{Ulrich Parlitz}
\author{Gerrit Wellecke}%
\email[]{gerrit.wellecke@ds.mpg.de}
\affiliation{Max Planck Institute for Dynamics and Self-Organization, Am Fa\ss{}berg 17, 37077 G\"ottingen, Germany}
\affiliation{Institute for the Dynamics of Complex Systems, University of G\"ottingen, Friedrich-Hund-Platz 1,  37077 G\"ottingen, Germany}

\keywords{
reservoir computing;
spatio-temporal chaos;
time series prediction;
echo state networks;
Kuramoto-Sivashinsky equation;
dimensionality reduction; 
machine learning;
recurrent neural networks
}

\date{\today}%

\begin{abstract}
Reservoir computers can be used to predict time series generated by spatio-temporal chaotic systems.
Using multiple reservoirs in parallel has shown improved performances for these predictions, by effectively reducing the input dimensionality of each reservoir. 
Similarly, one may further reduce the dimensionality of the input data by transforming to a lower-dimensional latent space.
Combining both approaches, we show that using dimensionality-reduced latent space predictions for parallel reservoir computing not only reduces computational costs, but also leads to better prediction results for small to medium reservoir sizes.
In the combined approach we further demonstrate that dimensionality reduction improves small--reservoir predictions regardless of noise contaminating the training data.
The benefit of 
dimensionality--reduced parallel reservoir computing is illustrated and evaluated on the basis of the prediction of the one--dimensional Kuramoto-Sivashinsky equation.

\end{abstract}

\maketitle
\section{Introduction}

Within recent years, reservoir computing \cite{jaeger_short_2001, maass_real-time_2002, verstraeten_experimental_2007} has been established as a computationally cheap machine learning method that leverages on driven dynamics of a high-dimensional dynamical system --- the reservoir --- to perform predictions. 
The reservoir itself is not trained, but subject to predefined reservoir properties. 
Within these constraints, the reservoir's structure is either initialised randomly in numerical implementations or determined by physical constraints in hardware implementations, referred to as physical reservoir computing \cite{tanaka_recent_2019,Rafayelyan2020}.
For training, a linear superposition of (functions of) the reservoir variables and the driving signals is optimised, usually by means of linear regression \cite{lukosevicius_reservoir_2009}.
Despite its simplicity and numerical efficiency, the reservoir approach is shown to perform well on sequential tasks such as time series prediction \cite{bianchi_other_2017, han_review_2021, bollt_explaining_2021, shahi_prediction_2022}.
However, the performance of the reservoir computing approach for the prediction of time series is often studied on trajectories of low-dimensional systems.

In practical applications, time series predictions are often required for high-dimensional systems, such as time series of spatio-temporal dynamics. 
The prediction of time series of high-dimensional dynamical systems, however, suffers from the so-called \textit{curse of dimensionality} \cite{bellman_dynamic_1957}. 
In the context of reservoir computing this means that very large reservoirs are required to enable accurate predictions.
This presents a problem, as large reservoirs demand increased computational run time and memory, especially during the training.
In addition, achieving long, accurate predictions requires careful hyperparameter tuning, which thus becomes particularly challenging and computationally expensive for  large reservoirs and memory requiring tasks \cite{fleddermann_enhancing_2025}.

For the prediction of spatio-temporal systems, the use of parallel reservoirs \cite{lu_reservoir_2017, pathak_using_2017, pathak_model-free_2018, zimmermann_observing_2018, herzog_reconstructing_2021, baur_predicting_2021, goldmann_learn_2022}, i.e. the splitting of the domain into multiple smaller subdomains, each predicted by its own reservoir, has been established as a method that enables reliable predictions of spatio-temporal systems with relatively small parallel reservoirs.
In addition to this method of reducing each reservoir's input dimension, latent space predictions \cite{liu_latent_2019, herzog_convolutional_2019, Ren2024, constante-amores_data-driven_2024, Lin_2021, Di_Antonio_et_al_2024, Mars-Gao_Kutz_2024} are a common data-driven method to effectively extract only relevant features of a high-dimensional data set, thereby often reducing the dimensionality of the data set. An overview of applications of dimensionality reduction in the data-driven modelling of spatiotemporal data can be found in Ref.~\cite{Pan_et_al_2023}.

In this paper, we combine the approach of parallel reservoirs with dimensionality-reduced latent space predictions and demonstrate improved performance for smaller, computationally cheaper reservoirs. 

We confirm that the use of parallel reservoirs increases prediction performance and, vice versa, serves as a well-functioning downsizing tool for the reservoir size. 
Moreover, we show that the combined approach of parallel dimensionality--reduced latent space predictions increases prediction performance for small reservoirs even further. 
This improvement is independent of the number of parallel reservoirs $\parres\geq2$, enabling reliable prediction performance with reduced computational cost.
Additionally, we consider the influence of noise in the training data. 
Artificial noise has been shown to stabilise iterative predictions \cite{wikner_stabilizing_2024}.

For small reservoirs, we find this behaviour similar to the parallel dimensionality-reduced method developed in this work.
In contrast to regularisation through noise, our novel approach does not suffer from deterioration of prediction performance for large reservoirs.
Further, if data is inherently noisy, e.\,g. due to experimental limitations, adding more noise may not yield any improvement.
We show that in this case our combined method is able to counteract noise and improve predictions.

The combined approach is presented and analysed based on iterative reservoir predictions of the one--dimensional Kuramoto-Sivashinsky equation (KSE) \cite{kuramoto_diffusion-induced_1978, sivashinsky_flame_1980} given by the partial differential equation (PDE)
\begin{equation}
\label{eq:ks}
    \partial_t u(x,t) = -\frac{1}{2} \partial_x \left[u^2(x,t)\right] - \partial_x^2 u(x,t) - \partial_x^4 u(x,t)\,,
\end{equation}
where $u$ is a spatio-temporal variable which evolves on a one--dimensional domain. 
Numerically integrated trajectories serve as ground truth, i.e. training and evaluation time series $u^{\,\mathrm{true}}(x,t)$ following \eq\eqref{eq:ks}. 
Details of the numerical procedure are summarised in Appendix~\ref{chap:numerics}.

Figure~\ref{fig:ks_prediction} displays the performance evaluation of an iterative prediction (see \sect\ref{chap:iterative_prediction}) for the KSE, by comparing a ground truth trajectory $u^{\,\mathrm{true}}(x,t)$ (Panel~\textbf{a}), to an iterative parallel dimensionality--reduced reservoir prediction $u(x,t)$ (Panel~\textbf{b}). 
\begin{figure}[t]
    \centering
    \includegraphics{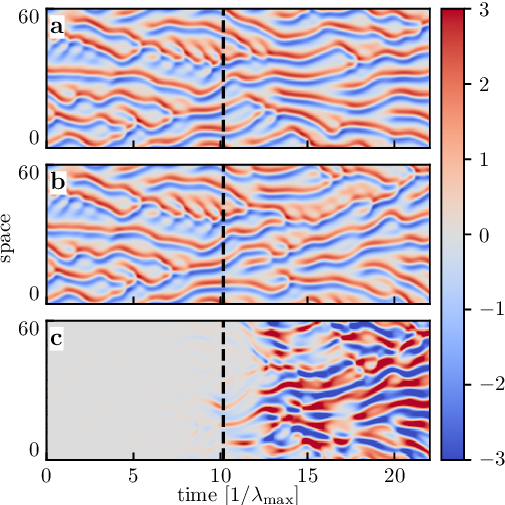}
    \caption{
        \textbf{Time series and iterative prediction of the Kuramoto--Sivashinsky model.}
        Panel \textbf{a} shows the temporal evolution of the trajectory following \eq\ref{eq:ks}, i.\,e. \emph{ground truth} data.
        In Panel \textbf{b}, the iterative prediction of the time series using the combined approach of parallel reservoirs with dimensionality reduction (see \sect\ref{chap:ext_lat_parres}) is shown.
        Panel \textbf{c} shows the difference between the ground truth and the prediction.
        The valid time of the prediction $\tvalid \approx 10$ Lyapunov times is marked by the dashed black line in all panels.
    }
    \label{fig:ks_prediction}
\end{figure}
The performed prediction shows a valid time (compare \eq\eqref{eq:validtime}) of $\tvalid\approx 10$ Lyapunov times (i.e. $\tvalid\approx 10 / \lambda_\mathrm{max}$ with $\lambda_\mathrm{max} \approx 0.095$ being the largest Lyapunov exponent, calculated with code from \cite{datseris_nonlinear_2022}).
This relatively long prediction horizon is slightly above achieved mean performances of $\tvalid\approx 9.2$ Lyapunov times and serves as an example of a very good prediction. 
The prediction was performed using the combination of parallel reservoirs with dimensionality reduction methods, introduced and analysed below,

and significantly exceeds typical valid times obtained using the classical reservoir computing approach:
Even with hyperparameter optimisation classical reservoir predictions with up to $\nodes\leq8000$ nodes achieve mean valid times below $\tvalid \leq 5$ Lyapunov times (compare \fig\ref{fig:ParRes} \textit{purple} or see \cite{Vlachas2020} for comparable results). 

Using parallel reservoirs, such predictions are possible given that the reservoirs are sufficiently large. 
Additional dimensionality reduction permits using significantly smaller reservoirs.

This manuscript first introduces the classical reservoir computing method and its application to iteratively predict time series in \sect\ref{chap:classicRC}. 
Subsequently, the parallel reservoir computing approach and its combination with latent space predictions is presented and analysed in \sect\ref{chap:ext_lat_parres}.
Lastly, we evaluate and discuss our findings with respect to their causes and the broader context in \sect\ref{chap:conclusion}.

\section{Reservoir Computing}
\label{chap:classicRC}
\subsection{Echo State Networks}
Following Jaeger~\etal \ \cite{jaeger2001,jaeger_optimization_2007}, we use time-discrete echo state networks as reservoirs, allowing for leaky integration.
The current state of the reservoir $\vec{s}_m$, at discrete time step $t_m=m\dt$, is given by 
\begin{equation}
\label{eq:inner_nodes_update}
    \vec{s}_m = (1-\alpha)\vec{s}_{m-1}  +\alpha \tanh\left(\nu \boldsymbol{W}_{\mathrm{in}}[b_{\mathrm{in}}, \vec{u}_m]\trans+\rho\boldsymbol{W}_{\mathrm{adj}}\vec{s}_{m-1}\right)\,,
\end{equation}
where $\inp=[u(\Delta x,t_m), u(2\Delta x,t_m), \ldots, u(D\Delta x,t_m)]\trans$ denotes the $D$-dimensional column vector of the time- and space-discrete driving signal and $\nu,\ \rho,\ \alpha$ are three hyperparameters scaling the input, spectral radius, and leaking rate, respectively. 
Further, $\win$ and $\wadj$ denote the input matrix and the adjacency matrix of the reservoir, respectively. 
The input matrix $\win$ maps the input vector $\iv=[\bin,\inp]\trans$ to the reservoir nodes (i.e. in a high-dimensional vector space $\real^\nodes$), where $[.,.]\trans$ denotes the concatenation of input bias $\bin$ and driving signal $\inp$ to a column vector.
The entries of $\win$ are independently drawn from a uniform random distribution of values in $[-0.5, 0.5)$.
The adjacency matrix $\wadj$ describes the inner connectivity of the reservoir. 
Its entries are drawn randomly from a uniform distribution of values in $[0,1)$.
However, only a fraction of all values is chosen from the distribution, as $\wadj$ is initialised as a random sparse matrix with an average degree $\degree$. 
In the last step of the initialization, the adjacency matrix is normalized by dividing all entries by the current spectral radius of the adjacency matrix, ensuring a spectral radius of one.

The reservoir states $\res$, the driving signal $\inp$, and an output bias $\bout$ are summarized in the extended state vector $\esv$ \cite{lukosevicius_reservoir_2009,pathak_hybrid_2018,shahi_machine-learning_2022, Parlitz2024}.
Following \cite{herteux_breaking_2020, pathak_model-free_2018, lu_reservoir_2017} we use the squared values of the second half of the reservoir states in the extended state vector. 
Thus, the extended state vector is given by $\esv = [s_{m,1}, \ldots, s_{m, N / 2}, s_{m,N / 2 + 1}^2, \ldots, s_{m,N}^2, \inp, \bout]\trans$.
In addition to the use of an input bias $\bin$, this is another common method to break symmetries of the reservoir dynamics \cite{herteux_breaking_2020}. 

The reservoir output $\outp=\wout\esv$ is obtained by linear superposition of the extended state vector's components. 
In the training data, for each input $\inp$ exists a desired reservoir output $\outp^\mathrm{true}$.
For the iterative prediction of time series, the desired output matches the next time step of the driving training time series $\outp^\mathrm{true}=\inp[m+1]^\mathrm{true}$.
The reservoir's output matrix $\wout$ is trained by minimising the regularised cost function $\sum_{m=1}^{\trainsteps}\|\outp^{\,\mathrm{true}}-\wout\esv\|^2+\beta\|\wout\|_2^2$ over $\trainsteps$ training time steps.
We summarise a time series in the extended state matrix $\esvmat = (\esv[1], \ldots, \esv[\trainsteps])\in \real^{(\nodes+\sysdim+1)\times\trainsteps}$ and the corresponding ground truth in a matrix $\outmat = (\outp[1]^{\,\mathrm{true}}, \ldots, \outp[\trainsteps]^{\,\mathrm{true}})\in \real^{\sysdim\times\trainsteps}$.
The global minimum of the cost function is given by
\begin{equation}
\label{eq:tikhonov}
    \boldsymbol{W}_{\mathrm{out}}=\boldsymbol{Y}\boldsymbol{X}\trans\left(\boldsymbol{X}\boldsymbol{X}\trans+\beta\boldsymbol{I}\right)^{-1} \,. 
\end{equation}
The regularisation parameter $ \beta $ disfavours large values in the output matrix.
This process is commonly referred to as \emph{Tikhonov regularisation} or \emph{ridge regression} \cite{hoerl_ridge_1970}.
Importantly, the computational cost of \eq\eqref{eq:tikhonov} increases as the number of nodes per reservoir $\nodes$ grows.

While the optimisation of the output matrix is straightforward, the performance of the reservoir computing approach strongly depends on the chosen hyperparameters.
A summary of the tested hyperparameters is shown in Table~\ref{tab:parameter}.
Within this work, we use a grid-search method to determine optimal values.
However, good performance is achieved only if the Echo-State-Property \cite{jaeger2001} is fulfilled, i.e. reservoir states are asymptotically uniquely determined by their driving sequence $(\vec{u}_m)_{m\in\mathbb{N}}$ and do not depend on their initialization $\vec{s}_0$, which is here chosen as uniformly random, $\vec{s}_0 \in [0,1)^N$.
To achieve convergence to the uniquely determined reservoir response, a transient or washout time $\ttrans$ is required.
During this time $\ttrans$, \eq\eqref{eq:inner_nodes_update} is computed while the reservoir state is not yet used for predictions.

\subsection{Iterative Time Series Predictions}
\label{chap:iterative_prediction}
Reservoir computers can be used to perform iterative predictions of chaotic time series by training a reservoir to predict the next time step, i.e. $\outp = \inp[m+1]$. 
The reservoir's output is then iteratively fed back to its input in a closed loop to predict the future evolution of the given time series. 
There are multiple measures of quality of such predictions.
Commonly used are normalised mean square errors averaged over many single--step predictions \cite{pathak_model-free_2018, Rafayelyan2020, Vlachas2020, Ren2024} or measurements of the replication of the attractor climate \cite{Lu2018}.
For the given study we measure the quality of an iterative prediction by measuring the \emph{valid time} $\tvalid$ defined as
\begin{align}
    \tvalid=\max_{E(t)<e}t, \quad \text{where}\quad
    E(t) = \frac{\|\vec{u}(t)-\vec{u}^{\,\mathrm{true}}(t)\|}{\langle\|\vec{u}^{\,\mathrm{true}}(t)\|^2\rangle_{\mathrm{t}}^{1/2}},
    \label{eq:validtime}    
\end{align}
where $e$ is a threshold value, that denotes the maximal accepted deviation between prediction and ground truth. 
Within this paper we consistently set $e=0.5$. 
Note that $E$ spatially averages the error on a discretised support such that it remains a function of time.
The valid time quantifies the ability of a reservoir to precisely predict a time series for as long as possible, knowing that due to the chaotic nature of the system, trajectories will diverge eventually.
To further generalise the chosen prediction measure, we rescale time by the largest Lyapunov exponent $\lambda_\mathrm{max} \approx 0.095$ of the KSE and use the Lyapunov time $1 /\lambda_\mathrm{max}$ as a meaningful system time scale. 
An example time series is shown in \fig\ref{fig:ks_prediction} to visualise the procedure.
The ground truth $\vec{u}^{\mathrm{true}}(t)$, shown in \fig\ref{fig:ks_prediction}~\textbf{a}, is integrated numerically following \eq\eqref{eq:ks} using a domain size of $L=60$ that is discretized with $D=128$ grid points (see Appendix~\ref{chap:numerics}).
Figure~\ref{fig:ks_prediction}~\textbf{b} shows the iterative prediction of the reservoir $\vec{u}(t)$ and \fig\ref{fig:ks_prediction}~\textbf{c} the deviation $\vec{u}(t)-\vec{u}^{\mathrm{true}}(t)$ between prediction and ground truth. 
Before generating the trajectory, the trained reservoir is run, without using its output, for a transient length of $\ttrans=25$ ($\approx 2.4$ Lyapunov times), which is omitted in the figure. 

At $t=0$ the iterative prediction starts and hence $\vec{u}(0)=\vec{u}^{\mathrm{true}}(0)$.
The error $E$ in  \eq\eqref{eq:validtime} exceeds the threshold $e=0.5$ at a valid time of  $t\approx 10$ Lyapunov times. 
The hyperparameters of the reservoir used in \fig~\ref{fig:ks_prediction} are summarized in \tab\ref{tab:parameter}. 
The reservoir is trained with $\trainsteps=50000$ training steps on a chaotic trajectory of the KSE of length $\ttrain=50000 \dt$ ($\approx1187.5$ Lyapunov times), where we use the sampling time $\dt = 0.25$. 

Throughout this work, we use the mean valid time of an optimised hyperparameter set as the measure of quality of different prediction approaches.
Therefore, for a given hyperparameter set, we average the performance over 10 randomly initialised reservoirs, each evaluated on 50 trajectories. 
The standard deviation between mean performances of the reservoirs, each averaged over 50 evaluation trajectories, serves as the uncertainty of the performance measure. 
Note that this neglects large performance fluctuations between different evaluation trajectories to isolate the performance fluctuations between different reservoir initializations.
Hyperparameters are optimised using a grid-search method.
Tested hyperparameter ranges are shown in \tab\ref{tab:parameter}.

\subsection{Spatio-temporal predictions require a large reservoir}
The large input dimensionality $\sysdim$ of spatio-temporal systems is a major problem of their prediction. 
This work considers a one--dimensional KSE system, however, this problem becomes even more evident, when considering higher--dimensional systems.
Similar discussions of this problem can be found in \cite{lukosevicius_practical_2012, zimmermann_observing_2018, Rafayelyan2020, baur_predicting_2021, shahi_machine-learning_2022}, relating poor performance of small reservoirs to the fact that ``the size of the reservoir must be large enough to provide rich dynamics and to capture the behaviour of the dynamical system represented by the input time series" \cite{shahi_machine-learning_2022}. 
Increasing the number of reservoir nodes seems to be necessary to achieve good reservoir prediction performance for spatio-temporal systems.
However, increasing the node number $\nodes$ significantly increases the run time (at least quadratically) and computational memory (linearly) of the reservoir training. 
Among others, increasing the number of reservoir nodes increases the size of the extended state matrix $\esvmat\in \real^{(N+\indim+1)\times\trainsteps}$.
This mainly contributes to the computational memory requirements and significantly prolongs the computation of \eq\eqref{eq:tikhonov}, as the involved matrices grow in size.
Finding means to reduce the size of well-performing reservoirs for the prediction of spatio-temporal systems is hence the primary objective of this study.

\section{Parallel Latent Space Predictions}
\label{chap:ext_lat_parres}
In the following, two concepts are presented to cope with the curse of dimensionality and high computational costs caused by large numbers of reservoir nodes. 
The first approach is based on a decomposition of the spatio-temporal dynamics into contiguous sub-domains, which are predicted in parallel by individual, relatively small reservoirs.
This approach is presented and analysed in Secs.~\ref{chap:ext_par_intro}-\ref{chap:ext_par_perform} and schematically depicted in  \fig\ref{fig:scheme_drrc}~\textbf{a}.
In this context, we discuss both the benefits and limitations of this method, which motivates the introduction of a second approach and allows for meaningful comparisons.
Therefore, we introduce (linear) dimensionality-reduced latent space transformations as another way to reduce the dimensionality of the reservoir's driving signal.
This method is depicted in \fig\ref{fig:scheme_drrc}~\textbf{b} and introduced in Secs.~\ref{chap:ext_lat_intro} and \ref{chap:ext_lat_choosevars}.
The combination of both approaches enables accurate predictions over long periods of time, despite using small reservoir systems, as demonstrated for the KSE in Sec.~\ref{chap:ext_drrc}.
Section~\ref{chap:noise_dr_similarity} explores how this dimensionality reduction is able to stabilise small--reservoir predictions, similar to artificial noise.
The combined method is also able to improve predictions if training data is subject to (substantial) amounts of noise, as demonstrated in Sec.~\ref{chap:noise_mitigation}.

\subsection{Parallel Reservoirs}
\label{chap:ext_par_intro}

\begin{figure*}[t]
    \centering
   \includegraphics[width=\textwidth]{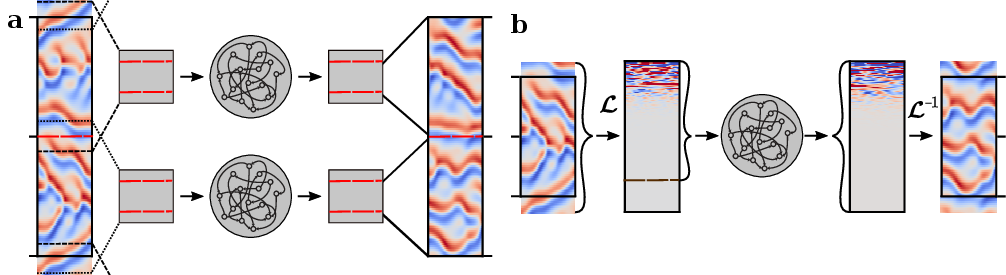}
    \caption{\textbf{Modifications of a single time series prediction step to enhance performance of reservoir computing.}
    \textbf{a}~The parallel reservoir approach is shown for $\parres=2$ parallel reservoirs. 
    The input domain is divided into two subdomains,  each predicted by its own reservoir. 
    Note that the input domains share overlapping neighbourhoods, while the prediction domains are disjoint.
    \textbf{b}~Dimensionality-reduced latent space predictions are shown using the PCA as linear transformation $\lintrafo$ of the system state. 
    In a second step, only the largest $\dimred=75\%$ of the PCA components are used as reservoir input.
    While the reservoir's input is only a portion of the transformed data, all transformed system variables are predicted.
    The inverse transformation $\lintrafo^{-1}$ maps the state back to the original space.}
    \label{fig:scheme_drrc}
\end{figure*}

The established approach to reduce the input dimensionality of a spatio-temporal system is the use of multiple reservoirs in parallel \cite{lu_reservoir_2017, pathak_using_2017, pathak_model-free_2018, zimmermann_observing_2018, herzog_reconstructing_2021, baur_predicting_2021, goldmann_learn_2022}.
The approach makes use of local states \cite{parlitz_prediction_2000}, i.e. the limited, spatial range of interactions in many physical systems.
In the case of the KSE, the temporal derivative $\partial_t u(t, x)$ at a fixed spatial coordinate $x\in[0,L]$ is solely determined by a local environment of the spatio-temporal variable $u(t, x)$ (see  \eq\ref{eq:ks}).
For sufficiently small time scales the system's dynamics are therefore spatially decoupled over sufficiently large distances. 
Hence, the domain can be split into several subdomains and single-step reservoir predictions, with sufficiently small prediction time steps, can be performed on each subdomain individually. 
In \fig\ref{fig:scheme_drrc}~\textbf{a} the approach of using $\parres=2$ reservoirs in parallel is sketched for predictions of the one--dimensional KSE.

The subdomain, predicted by an individual reservoir, is called the core $\corevar\in\real^{\coredim}$ of the domain of the $\parind$-th reservoir. 
Interactions between subdomains are included by adding the surrounding of each core --- the neighbourhood $\neivar\in\real^{\neidim}$ --- to the reservoir's input vector, i.e. $\ivpar = [\bin, \corevar, \neivar]\trans\in \real^{1+\indim}$, with input dimensionality $\indim =\coredim+\neidim$, where the indices c, n correspond to the core and neighbourhood, respectively.
For iterative time series predictions, each reservoir is trained to predict the next time step of its core variables, $\outppar = \corevar[m+1]$.
In each prediction time step, first all parallel reservoirs perform individual predictions. 
Then, the whole state of the predicted system $\inp[m+1]=[\corevar[m+1][1], \ldots, \corevar[m+1][\parres]]\trans$ is merged together by combining all predicted cores. 
Thereby, the input of each reservoir, including core and neighbourhood, is updated with predictions of itself and adjacent reservoirs.
The number of parallel reservoirs $\parres$ and the physical length of the neighbourhood $\neilen = \neiind\Delta x$, which is an integer multiple $\neiind$ of the spatial discretisation $\Delta x$, are two additional hyperparameters that determine the input dimensionality $\indim$ of each parallel reservoir. 
Recall that in this work we use a one--dimensional domain $[0,60]$ with periodic boundary conditions discretized with $D=128$ grid points.
However, the introduced methods generalise to $\spacedim$-dimensional cubes for system and core domains, where $\spacedim$ is the dimensionality of the domain of the spatio-temporal system.
For a system with a total number of $\sysdim$ grid points (combining all spatial dimensions), hence the dimensions of core, neighbourhood, and input are given by
\begin{align}
    \coredim &= \sysdim/M,\\
    \indim &= (2\neiind+{\sqrt[\spacedim]{\coredim}})^\spacedim,
    \label{eq:parresdim}\\
    \quad\neidim &= \indim -\coredim,
\end{align}
respectively. 

After discussing equivalent formulations and the computational gain of the proposed method, we analyse the performance of the parallel reservoir computing approach with respect to the two parameters $\parres$ and $\neiind$ for predictions of the one--dimensional KSE \eqref{eq:ks} in \sect\ref{chap:ext_par_perform}.

\subsection{Physics-Informed Weight Matrices, Translational Invariance, and Computational Gain}
\label{chap:ext_par_theo}
Theoretically the use of $\parres$ parallel reservoirs with $\nodes$ nodes each, is equivalent to using a large reservoir of $\parres\nodes$ nodes with predefined structures of input matrix $\win$, adjacency matrix $\wadj$, and output matrix $\wout$. 
In this case, the predefined structure of weight matrices incorporates physical knowledge of the local nature of the PDE (see Appendix ~\ref{app:physics_informed_weightmatrices}). 

Parallel reservoirs (instead of pre-structured weight matrices) are used in the prediction of spatio-temporal systems due to the simplicity of their implementation and the computational efficiency, as parallel reservoirs allow for sequential or parallel training of reservoirs and may benefit from translational invariance of the dynamics.
This can greatly reduce computational costs of handling large reservoirs or input systems.
If only one large reservoir with $\parres\nodes$ nodes is used, the training is significantly more memory intensive compared to the prediction or transient phase.
This is due to the need for storing and performing computations (compare \eq\eqref{eq:tikhonov}) with the extended state matrix $\esvmat$.
Using a single reservoir with $\parres\nodes$ nodes on the whole input domain it is $\esvmat\in\real^{(\parres(\nodes+\coredim)+1)\times\trainsteps}$, where usually the dimensionality of the matrix in temporal direction is much larger, i.e. $\trainsteps \gg \parres(\nodes+\coredim)+1$.
By using parallel reservoirs the input dimensionality is reduced from $\parres\coredim$ to $\indim = \coredim + \neidim$ and the node number $\nodes$ by a factor $\parres$. 
This greatly reduces the memory requirements during training.

In case of dynamical systems with translational symmetry, such as the KSE, the computational advantages are even greater. 
Here, the dynamics in each subdomain follow identical rules, i.e. the same homogeneous differential equation without spatial dependencies. 
Therefore, it suffices to train a single reservoir which is duplicated and applied to each subdomain. 
This method has been applied and demonstrated by several previous works \cite{zimmermann_observing_2018, goldmann_learn_2022, barbosa_learning_2022}.
Depending on the amount of available data, the training data for this reservoir can optionally consist of the data of one single subdomain or be a combination of all the subdomains. 
In the latter case, successively through all subdomains, the reservoir is first propagated on a transient before the temporal evolution of the reservoir states and driving signals are recorded into the extended state matrix $\boldsymbol{X}$. 
Similarly, the desired reservoir outputs are concatenated in the output matrix $\boldsymbol{Y}$.
Thereby the matrices $\boldsymbol{X}$ and  $\boldsymbol{Y}$ consist of training data from all subdomains.
After training, the reservoir is duplicated, such that $\parres$ different reservoir states $\res^{(i)}$, with $i\in\{1,...,\parres\}$, exist in parallel --- one for each subdomain.
The training of only a single parallel reservoir computer drastically reduces the computation time of the memory-intensive training period.
On the contrary, computational demands (i.e. number of operations) during evaluation (i.e. transient and prediction steps), do not benefit from homogeneous systems. 
However, using a single set of weight matrices ($\winpar, \wadjpar, \woutpar$) for all parallel reservoirs requires less memory.

\subsection{Performance of Parallel Reservoirs}
\label{chap:ext_par_perform}

\begin{figure}[t]
\includegraphics[scale=1.0]{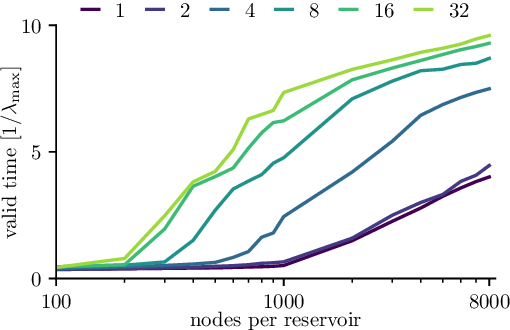}
    \caption{\textbf{Parallel reservoirs improve prediction performance for fixed node numbers.}
    The mean valid time is shown for varying reservoir sizes $\nodes$ (logarithmic scale on the $x$-axis) and parallel reservoirs $\parres$ (\textit{coloured lines}). 
    Optimal hyperparameters are determined for each case individually and mean performance is averaged over 500 predictions, obtained from 10 random reservoir realisations, each evaluated on 50 trajectories.
    }
    \label{fig:ParRes}
\end{figure}

We evaluate the performance of the parallel reservoir approach based on iterative time series predictions of the one--dimensional KSE (see \eq\eqref{eq:ks}) of length $\spacelen=60$. 
Figure \ref{fig:ParRes} demonstrates the performance gains due to increasing numbers $\parres$ of parallel reservoirs for a fixed neighbourhood dimensionality $\neidim=2\cdot10$, i.e. adding a spatial domain of length $ \neilen = 10\Delta x$ in each direction of all prediction cores.

The mean performance of reservoirs with optimised hyperparameters (see \tab\ref{tab:parameter}) improves with increasing numbers of parallel reservoirs. 
However, varying the number of parallel reservoirs from $\parres=1$ to $\parres=2$ has almost no effect on the performance. 
Great performance increases are achieved varying $\parres = 2$ to $\parres=4$ and from $\parres = 4$ to $\parres = 8$. 
Only slight performance increases can be achieved by increasing $\parres$ even further. 
Note that the diminishing performance increase is consistent with the diminishing reductions of input dimensionality for increasing numbers of parallel reservoirs, $\indim = 128/\parres + \neidim\xrightarrow{M\gg 1}\neidim$.
Nonetheless, increasing the number of parallel reservoirs generally improves performance.

\begin{table}[ht]
  \caption{
    Tested ranges or values of parameters that are used in the hyperparameter optimisation of time series predictions of the KSE, including  classical hyperparameters, parameters attributed to parallel reservoirs and to latent space predictions.
    Classical hyperparameters (\textit{top}), are always optimised. 
    In \sect\ref{chap:ext_par_perform} additionally the hyperparameters attributed to parallel reservoirs (\textit{middle}) are varied. 
    In \sect\ref{chap:ext_drrc} all parameters are varied.
  }  %
    \begin{tabular}{llcc}
        \toprule
        \multicolumn{2}{c}{Hyperparameter}         & Tested Values             & Figure \ref{fig:ks_prediction} \\
        \midrule
        $\spec$            &spectral radius         & $[10^{-2}, 10]$             & $3.162278$           \\
        $\inscale$         &input scaling           & $[10^{-4}, 10]$             & $1.7783$           \\
        $\degree$          &adjacency degree        & $2,3$         & $2$         \\
        $\alpha$           &leaking rate            & $0.9, 1$           & $0.9$         \\
        $\dt$              & sampling time               & $0.25$          & $0.25$        \\ 
        $\beta$            &regularisation const.   & $[10^{-6}, 10^{-2}]$  & $10^{-6}$     \\
        $\nodes$           &reservoir nodes         & $[100, 8000]$           & $8000$        \\
        \midrule
        $\parres$          & parallel reservoirs    & $2^0, 2^1, ..., 2^5$  & $8$           \\
        $\neilen$          & neighbourhood length    & $[2\Delta x, 10\Delta x]$    & $10 \Delta x$ \\
        \midrule
                           & transformation                 & FFT, PCA              & PCA\\
        $\dimred$          & dim. reduction [\%]        & $25, 50, 75, 100$ & $50$\\
        \bottomrule
    \end{tabular}
    \label{tab:parameter}
\end{table}

\begin{figure}[ht]
\includegraphics[scale=1.0]{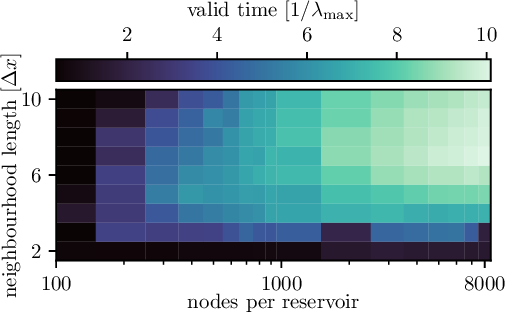}
    \caption{\textbf{Optimal neighbourhood length exists and is dependent on the reservoir size.}
    The mean valid time is shown for a given number of nodes $\nodes$ and neighbourhood length $\neilen$ for $\parres=32$ parallel reservoirs (see appendix \fig\ref{fig:ParRes_nei_all} for qualitatively similar results of other numbers of parallel reservoirs).
    Best-performing neighbourhood lengths for $\nodes>200$ nodes are in $[5\Delta x, 8\Delta x]$. 
    Valid time declines sharply when the neighbourhood length is smaller than the optimal value.
    }
    \label{fig:ParRes_nei}
\end{figure}

While more parallel reservoirs consistently increase prediction performance, an optimal neighbourhood length $\neilen$ exists. 
Figure \ref{fig:ParRes_nei} shows mean valid times of $\parres=32$ parallel reservoirs with optimised hyperparameters for different neighbourhood lengths $l \in [\Delta x, 10 \Delta x]$ and node numbers $\nodes\in[100,8000]$.

\begin{figure}
    \centering
    \includegraphics[scale=1]{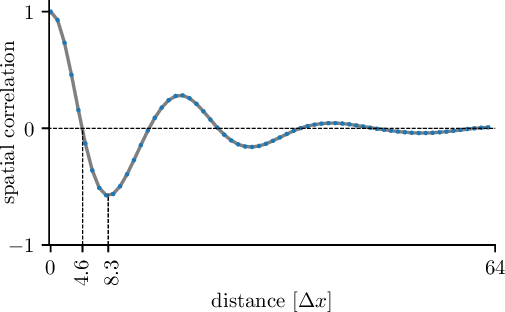}
    \caption{
      \textbf{Autocorrelation function of the KSE has large magnitudes for short distances and decays towards zero for large distances.}
      The wave-like spatial structure of the system (compare \fig\ref{fig:ks_prediction}), induces alternations between high positive and negative values of spatial correlation.
      The first zero crossing is at a distance of $\approx 4.6\Delta x$ and the first minimum at $\approx 8.3\Delta x$.
    }
    \label{fig:correlation}
\end{figure}

For each given node number, a best-performing neighbourhood length exists whose value slightly increases with increasing reservoir size. 
Best-performing neighbourhood lengths for $\nodes>200$ nodes are in $[5\Delta x, 8\Delta x]$. 
The optimal neighbourhood length can be compared with the spatial correlation of the system, which is illustrated in \fig\ref{fig:correlation}.
The spatial wave-like patterns of the KSE result in decaying oscillations of the spatial correlation function.
The best-performing neighbourhood length agrees with the order of magnitude between the first zero crossing (at~$\approx4.6\Delta x$) and the minimum (at~$\approx 8.3 \Delta x$) of the systems spatial correlation.

Qualitatively similar behaviour, with best-performing neighbourhood length in $[3\Delta x,8\Delta x]$ for $\nodes=8000$, is obtained for other numbers of parallel reservoirs and is shown in the appendix (see \fig\ref{fig:ParRes_nei_all}).
Overall, these results indicate the need of sufficiently large neighbourhoods for accurate reservoir predictions but also the existence of an optimal neighbourhood size, as the neighbourhood increases dimensionality of the input.
Since the performance decrease from the best performing neighbourhood length is steeper towards smaller neighbourhoods, we use a neighbourhood length of $\neilen = 10\Delta x$ within the following. 
While this choice is non-optimal, i.e. better prediction performance is achieved with smaller neighbourhood length, qualitative results are independent from this choice (compare appendix \fig\ref{fig:drrc_nei_compare}).

The use of parallel reservoirs offers a computationally feasible approach to tackle challenges of predicting (high-dimensional) spatio-temporal systems. 
However, its potential as a down-sizing tool for reservoirs is limited, as the input size can not be reduced indefinitely and thus performance gains diminish with an increasing number of parallel reservoirs.
Therefore, an additional method is presented in the following.

\subsection{Latent Space Predictions}
\label{chap:ext_lat_intro}
Dynamical systems often exhibit dynamics constrained to a lower-dimensional subset, such as a strange attractor, within the high-dimensional state space. 
Moreover, the variables that describe the system may not provide the clearest view on its intrinsic dynamics. 
In the field of machine learning, a common approach is to use a transformation that maps observed data into a, usually lower-dimensional, latent space, where the essential dynamical features become more accessible \cite{liu_latent_2019, herzog_convolutional_2019, racca_predicting_2023, shahi_machine-learning_2022, Ren2024}. 
Latent space predictions have been used to enable or enhance the prediction of spatio-temporal systems \cite{herzog_reconstructing_2021, Ren2024, constante-amores_data-driven_2024}. 
Further they are explored to improve reservoir computer predictions by extracting essential features temporally from an univariate time series \cite{shahi_machine-learning_2022} or spatially from spatio-temporal time series \cite{racca_predicting_2023}.
In spatio-temporal systems high redundancy of information is given by large spatial cross-correlation in local neighbourhoods (compare \fig~\ref{fig:correlation}).

Therefore, the approach of parallel reservoirs is commonly paired with a dimensionality reduction approach of zero-th order \cite{parlitz_prediction_2000, zimmermann_observing_2018, herzog_reconstructing_2021}, which can be understood as a latent space representation of the subdomain.
That is, in addition to the partitioning of the domain into subdomains, local redundancies are removed from each subdomain by subsampling the spatial variable, i.e. considering only every $\measspace$-th grid point in each spatial direction. 
Without an in-depth analysis of performance dependence on the subsampling spacing $\measspace$, the presented approaches are shown to be effective in time series and cross-predictions of spatio-temporal systems \cite{zimmermann_observing_2018, herzog_reconstructing_2021}. 
While the presented approaches deliver promising results, we suggest the use of higher-order transformations to test the use of (parallel) latent state predictions for spatio-temporal systems. 
As a first step, we use well-known linear, i.e. first-order, transformations $\lintrafo$ namely principal component analysis (PCA) or fast Fourier transformation (FFT) to transform and thereafter reduce the high-dimensional spatially discretized input $\inp^{(i)}\in\real^\indim$ of each parallel reservoir. 
However, the presented and implemented framework is in principle applicable to arbitrary (non-linear) transformations for which an inverse mapping $\lintrafo^{-1}$ is defined. Combinations of linear dimensionality reduction based on FFT or PCA and reservoir computing have previously also been  proposed to forecast geophysical systems  \cite{Lin_2021} and for developing digital twins of brain dynamics \cite{Di_Antonio_et_al_2024}.

Figure~\ref{fig:scheme_drrc}~\textbf{b} schematically shows one time step of a latent state prediction, supplemented with dimensionality reduction. 
To visualise the dynamic evolution of the state, not only one time step, a time series of states is shown.
In the scheme, the system state $\inp$ (\textit{left}) is transformed with the PCA as linear transformation $\lintrafo$. 
The decay of amplitude with increasing principal component index (top to bottom) is clearly visible in the transformed domain (\textit{second from left}). 
Only a fraction of $\dimred = 75\%$  of the principal components are used in the input vector $\iv$ of the reservoirs.
Still, the full vector $\lintrafo \inp[m+1]$ of principal components (\textit{second from right}) is trained to be predicted by the reservoir to allow the application of the inverse transformation $\lintrafo^{-1}$. 
In a last step, the inverse transformation $\lintrafo^{-1}$ is applied to the predicted output, to restore the next time step of the time series $\inp[m+1]$ (\textit{right}), thus closing the loop in iterative applications. 
While time series predictions with a single reservoir can be iterated directly in latent space, i.e. without using the inverse mapping $\lintrafo^{-1}$ in each time step, the presented framework generalises to multiple parallel reservoirs (see \sect\ref{chap:ext_lat_choosevars}), where the synthesis of predictions is required in real space.

\subsection{Choosing Relevant Latent Space Variables}
\label{chap:ext_lat_choosevars}
The linear transformations are supplemented with dimensionality reduction, such that only a fraction $\dimred$ of the FFT modes or principal components are used as reservoir input.
To easily generalise the approach to latent space predictions with arbitrary transformations $\lintrafo:\real^\indim\to\real^\indim$, we suggest the following procedure:
\begin{enumerate}
    \item Sort transformed variables $\lintrafo\inp$ in decreasing order of relevance using a permutation matrix $\order$ --- we will give meaning to what `relevance' means later on.
    \item Include only  sufficiently relevant latent state variables in the reservoir's input $\iv=[\bin, \prodr\order\lintrafo\inp]\trans$, where $\prodr$ is a projection on the first $\dimred \indim$ variables.
\end{enumerate}
The ordering of PCA modes is trivial, since ordering is part of the trained PCA.
Here, the amplitude of the principal component, which serves as a good measure of the relevance of the component, decays with its index (compare \fig\ref{fig:scheme_drrc}). 
We hence propose an identity transformation as ordering permutation, and therefore using the first $\dimred\indim$ principal components as the reservoirs input. 
For the FFT the selection of relevant modes is not trivial.
Here, we propose to order the spatial FFT modes with decreasing temporally maximal amplitude. 
That is, for the vector of temporal maxima of FFT modes $\vec{v}=\max_{m\leq\trainsteps}|\lintrafo\inp|$, we define an permutation $\sigma$, such that $v_{\sigma(1)}\geq v_{\sigma(2)}\geq\ldots\geq v_{\sigma(\indim)}$ and use the corresponding permutation matrix
\begin{align}
    \order_{ij} = \begin{cases}
1, & \text{if } j = \sigma(i), \\
0, & \text{else},
\end{cases}
\end{align}
to order the FFT modes.
However, this choice is somewhat ambiguous, and different measures of relevance, such as largest temporal variance, are good alternative choices and provide similar results.

\begin{figure*}[t]
  \includegraphics[width=\textwidth]{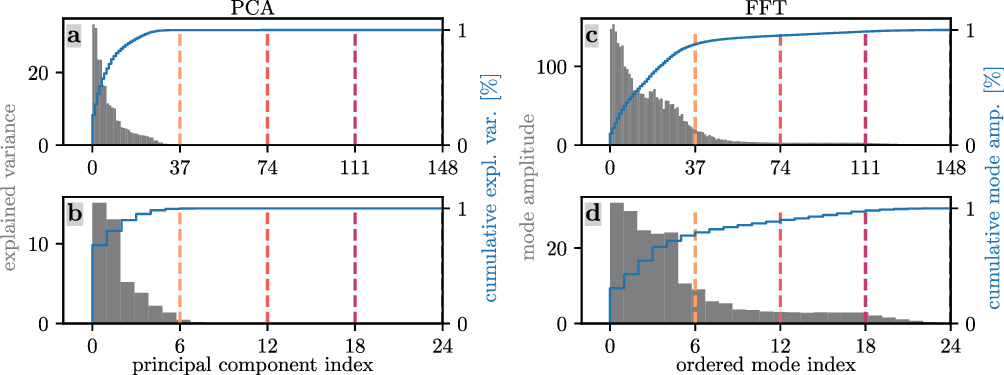}
  \caption{\textbf{Dominant modes contain the relevant information for time series prediction.}
    (\textbf{a},\textbf{b}) Principal components of a time series of the KSE.
    Here, the relative cumulative explained variance shows that almost no variance is explained along $75\%$ of all principal components.
    (\textbf{c},\textbf{d}) Ordered FFT modes for a time series of the KSE. 
    Similar to the PCA, an FFT of the given system also concentrates most of the information in the dominant half of the modes.
    Note that panels (\textbf{a},\textbf{c}) and (\textbf{b},\textbf{d}) correspond to $\parres=1$ and $\parres=32$ parallel reservoirs, respectively.
  }
  \label{fig:latentdr}
\end{figure*}

The decay of chosen measures of relevance with increasing index of ordered latent space variables are depicted in \fig\ref{fig:latentdr}. 
Panels~\textbf{a} and \textbf{c} show a monotonic decrease of explained variances with increasing principal component index (\textit{grey}) for $\parres=1$ and $\parres=32$ parallel reservoirs, respectively. 
Principal component indices that constitute $100\%, 75\%, 50\%$ and $25\%$ of all components are marked with dashed lines in \textit{violet}, \textit{pink}, \textit{dark orange} and \textit{light orange}, respectively. 
The cumulative explained variance, shown in \textit{blue} as fraction of the total cumulative explained variance, reaches values close to one already at $25\%$ of all principal components.
Similar results are shown for ordered FFT modes in \fig\ref{fig:latentdr}~\textbf{c} and \textbf{d} for $\parres=1$ and $\parres=32$ parallel reservoirs, respectively. 
However, for the FFT the decay of amplitude with increasing ordered mode index is not monotonous. 
The deviations from a monotonous distribution result from choosing the ordering $\order$, based on $\ressamples=10$ training data sets and calculating the depicted distribution based on temporal maximal values of $\order\lintrafo\inp$ over only one training data set. 
This highlights the sensitivity of the selected ordering of FFT modes to the amount of training data, reflecting the sensitivity of the maximum to outliers, i.e. modes with high amplitude for short time. 
Note that here sensitive dependence on outliers is not a bug, but a relevant feature of the chosen ordering $\order$. 
A less sensitive condition (such as the ordering with decreasing temporal mean) has been tested with worse prediction performance, indicating that some FFT modes which are relevant for good predictions are rarely excited with large amplitude.

If latent state predictions are combined with parallel reservoirs, the driving signal of each reservoir $\inppar$, with $\parind\leq\parres$ as the index of the parallel reservoir, is transformed using the transformation $\lintrafo$ and its dimensionality is reduced through $\prodr\order$. 
The input vector of each reservoir is hence given by $\iv^{\,(\mathrm{\parind})} = [\bin, \prodr\order\lintrafo\inp^{\,(\mathrm{\parind})}]\trans$. 
Each reservoir is trained to predict all transformed variables of its input domain $\outp^{\,(\mathrm{\parind})} = \lintrafo\inp[m+1]^{\,(\mathrm{\parind})}$. 
The inverse transformation $\lintrafo^{-1}$ restores the whole input domain, including core and neighbourhood cells.
However, it can be assumed, that predictions on neighbourhood cells are not accurate, due to the influence of unknown neighbouring cells.
The whole state vector of the next time step is synthesised by combining the core cells $\inp[m+1] = [\proc\lintrafo^{-1}\outppar[m][1], \ldots, \proc\lintrafo^{-1}\outppar[m][\parres]]\trans$, neglecting the flawed predictions of neighbourhood cells. 
This approach ensures that the reservoir does not have to predict the inverse transformation.

\subsection{Performance of Parallel Latent Space Predictions}
\label{chap:ext_drrc}

\begin{figure}[ht]
\includegraphics[scale=1.0]{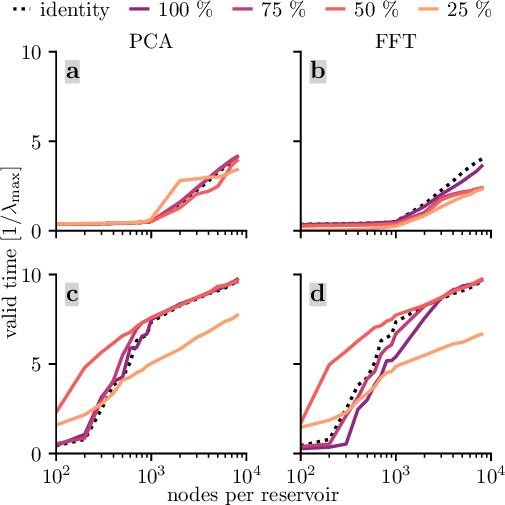}
\caption{
        \textbf{Linear dimensionality reduction improves prediction performance of small parallel reservoirs.}
        Panels show mean prediction performance using PCA (\textit{left}) or FFT (\textit{right}).
        For comparison, the mean prediction performance without dimensionality reduction is shown in all panels as a black dashed line, labelled \emph{identity}.
        With $\parres=1$ reservoir (\textbf{a} and \textbf{b}) dimensionality reduction most often decreases mean prediction performance.
        With $\parres=32$ parallel reservoirs (\textbf{c} and \textbf{d}), dimensionality reduction to $\dimred=25\%$ or $\dimred=50\%$ improves performance for small reservoirs $\nodes<1000$.
        For large node numbers, reduction to $25\%$ worsens prediction performance.  
        Results for other $\parres\geq2$ are qualitatively similar (compare appendix \fig\ref{fig:drrc_nei_compare}).
    }
    \label{fig:LatentParRes}
\end{figure}

Within this section we analyse the performance of parallel latent space predictions as previously illustrated. 
We use $\parres\in \{1,32\}$ parallel reservoirs and employ PCA and FFT to reduce the input dimensionality to $\dimred\in \{100\%, 75\%, 50\%, 25\%\}$, see \fig\ref{fig:LatentParRes}.
Note, however, that the results displayed here qualitatively generalise for any number of parallel reservoirs $\parres \ge 2$, see also Appendix~\ref{sec:Generality_of_Qualitative_Results}.
Generally the combined approach is compared to the case of no dimensionality reduction, which we coin \emph{identity}.

Using a single reservoir, $\parres = 1$, significant increases in performance compared to the untransformed case are observed only for a reservoir with $\nodes=2000$ nodes when using $\dimred=25\%$ of the principal components, see \fig\ref{fig:LatentParRes}~\textbf{a}. 
In all other cases, performance is either similar or worse, compared to predictions without linear transformation and dimensionality reduction.
Note, that especially iterative predictions of subsets of the FFT modes (see \fig\ref{fig:LatentParRes}~\textbf{b}), i.e. $\dimred\leq100\%$, significantly worsen predictions compared to the case of untransformed parallel predictions.

Using $M = 32$ parallel reservoirs combined with linear dimensionality reduction, mean predictions are improved, see \fig\ref{fig:LatentParRes} \textbf{c},\textbf{d}.
Here, the input dimensionality of each reservoir is already reduced to $\indim = 4+20$ by using the parallel reservoir approach (compare \eq\ref{eq:parresdim} with $\neiind=10$ neighbourhood cells in each direction and a system dimensionality $d=1$).
For the PCA (see \fig\ref{fig:LatentParRes}~\textbf{c}), slight performance improvements for arbitrary reservoir sizes are observed using $\dimred = 100\%$ (\textit{violet}) and $\dimred = 75\%$ (\textit{pink}) of all $\indim$ principal components.
Reducing the reservoir's input to only $\dimred = 50\%$ (\textit{dark orange}), i.e. using only $\dimred\indim=12$ input dimensions, significantly increases the performance for reservoirs with up to $\nodes = 1000$ nodes and leads to slight performance gains for even larger reservoirs. 
Decreasing input dimensionality to $\dimred=25\%$ (\textit{light orange}), leads to even greater performance gains for small reservoirs (up to $\nodes=200$), while decreasing the performance for large reservoirs ($\nodes > 1000$). 
Qualitatively similar results for substantial dimensionality reduction (to $\dimred\leq50\%$ of the input dimensions) are shown in \fig\ref{fig:ParRes}~\textbf{b} and \textbf{d} using maximal FFT modes and $\parres=1$ and $32$ parallel reservoirs, respectively. 
In contrast to slight performance gains observed for predictions with $\dimred = 100\%$ and $\dimred = 75\%$ of the principal components, slight performance losses are shown for these values of $\dimred$ using maximal FFT modes.
Notably, we observe a difference in the performance between predictions using the untransformed input (\textit{identity}), using $100\%$ of principal components or using $\dimred=100\%$ of FFT modes.
However, these three cases represent the same (local) system state expressed in different bases.
This highlights that different representations of the (local) state, i.e. different ways of encoding the system's dynamical features, cause a change in the capabilities of reservoir computers to effectively process the provided information.

In summary, combining parallel reservoirs with dimensionality reduction can significantly enhance prediction performance. 
This enables the use of computationally cheap predictions of small reservoirs with less than $\nodes=500$ nodes in parallel latent space predictions that outperform huge reservoirs with $\nodes\geq8000$ nodes in the classical reservoir application. 
To ensure a sufficiently large neighbourhood dimensionality, we choose $\neidim = 2 \cdot 10$. 
Recall that predictions suffer more strongly from a too small neighbourhood, making it a good rule of thumb to slightly overestimate this parameter.
See \sect\ref{chap:ext_par_perform} and Appendix~\ref{sec:Generality_of_Qualitative_Results} for a more detailed analysis of this parameter, showing that different parameter choices yield the same qualitative results.
Further, the result of improved performance for small reservoir sizes generalises for arbitrary numbers $\parres\geq2$ of parallel reservoirs, which we elaborate in Appendix \ref{sec:Generality_of_Qualitative_Results}.

\begin{figure}[t]
\includegraphics[scale=1.0]{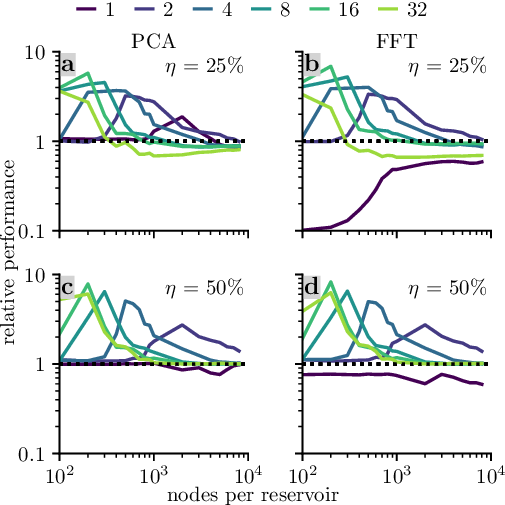}
\caption{
        \textbf{Dimensionality reduction consistently improves predictions of small parallel reservoirs.}
        The relative mean performance, i.e. the ratio of performance with and without dimensionality--reduced latent space transformation, is shown for various numbers $\parres$ (\emph{colours}) and sizes $\nodes$ of parallel reservoirs.
        Black dashed lines facilitate the identification of performance increases, $> 1$, and decreases $< 1$.
        (\emph{top}) Strong dimensionality reduction, $\dimred = 25\%$, with PCA (\textbf{a}) and FFT (\textbf{b}), significantly improves small--reservoir predictions, but degrades performance of large reservoirs.
        (\emph{bottom}) Intermediate dimensionality reduction, $\dimred = 50\%$, with PCA (\textbf{c}) and FFT (\textbf{d}), generally improves performance for parallel reservoirs $\parres \ge 2$. 
        In the case of large reservoirs, $\nodes > 1000$, dimensionality--reduced prediction performance converges to that of the classical parallel approach.
        Compare with \fig\ref{fig:LatentParRes} for coressponding absolute performances.
    }
    \label{fig:LatentParRes_rel}
\end{figure}
Figure~\ref{fig:LatentParRes_rel} shows the relative performance \mbox{$\fper(\parres, \nodes) = \tvalid(\parres, \nodes)/\tvalid'(\parres, \nodes)$} for different numbers of parallel reservoirs $\parres$ over different reservoir sizes $\nodes$.
Here $\tvalid(\parres, \nodes)$ denotes the valid time of parallel latent space predictions with dimensionality reduction and $\tvalid'(\parres, \nodes)$ without transformation and dimensionality reduction.
The relative performance simplifies the evaluation of parallel latent space predictions as values $\fper>1$ indicate improvement and $\fper<1$ decline of predictions using dimensionality reduction.

Significant increase in performance, $\fper>1$, is shown for small parallel reservoirs across both transformations, all numbers of parallel reservoirs, $\parres\geq2$, and dimensionality reduction fractions, $\dimred\leq50\%$.
This underlines the generality of the performance increase by dimensionality reduction for reservoirs that are otherwise too small to extract relevant features of high-dimensional input data, which often leads to unstable iterative predictions.
As the number of parallel reservoirs increases, the reservoir size $\nodes^*$, at which the highest relative performance $\fper(\parres,\nodes^*)$ is achieved, shifts towards smaller values. 
This can be explained by the independence of the two previously discussed effects: dimensionality reduction enhances the performance of smaller reservoirs (see \fig\ref{fig:LatentParRes}), while increasing the number of parallel reservoirs also improves performance for smaller reservoir sizes (see \fig\ref{fig:ParRes}).
The largest relative performance is therefore observed at node numbers $N^*$ where classical parallel predictions still fail, i.\,e. $\tvalid \approx 0$, but the combined approach of parallel latent space predictions achieve substantial valid times, $\tvalid \gg 0$.
Consequently, similar to shifts in high absolute performance by increasing the number of parallel reservoirs, the maximum of the relative performance $\fper(\parres,\nodes^*)$ shifts towards smaller reservoirs (see \fig\ref{fig:LatentParRes_rel}).

In addition to shifts, the magnitude of highest relative performance $\fper(\parres, \nodes^*)$ mostly grows with increasing numbers of parallel reservoirs, showing that latent space predictions work well, not despite, but rather because of using parallel reservoirs.
This reflects that the different dimensionality reduction methods, i.e. using local and latent space predictions, leverage independent characteristics of spatio-temporal data.
That is, they reduce the input dimension, firstly by enforcing decoupled reservoir dynamics which makes use of decoupled spatio-temporal dynamics and, secondly, by utilising low-dimensional latent space representations of the local state, effectively removing local redundancies. 
The outliers to the trend of increased relative performance with increasing number of parallel reservoirs are given by $\parres=32$ parallel reservoirs and might be attributed to low resolution of the number of nodes per reservoir in the relevant region (for nodes in $\nodes\in[100,200]$).

For both considered transformations and high numbers of parallel ($\parres\geq8$), large reservoirs ($N>1000$), relative performance is below $\fper=1$ if input dimensionality is reduced to  $\dimred=25\%$ (see \fig\ref{fig:LatentParRes_rel}~\textbf{a} and \textbf{b}) and saturates towards  $\fper=1$ if input dimensionality is reduced to $\dimred=50\%$.
This shows that for large reservoirs, which can efficiently process high-dimensional input data, dimensionality reduction reduces performance if too many variables are neglected ($\dimred=25\%$) and has no influence on the performance if the dimensionality is reduced to a proper amount ($\dimred=50\%$). 
Comparing with \fig\ref{fig:latentdr}~\textbf{b}, we see that for $\dimred=25\%$ principal components which visibly explain a non-zero variance are removed from the input data set, while all components which are neglected for $\dimred=50\%$ explain a variance $\ll1$.
Outstanding losses in performance ($\fper<1$) are observed for predictions using maximal FFT modes on the whole domain, i.e. $\parres=1$. 
As this is only visible for FFT modes, it might be attributed to the chosen method of mode selection.
We conjecture, that for large input domains (of one reservoir) the choice of selected FFT modes suffers from fine resolution of maxima within the frequency spectrum, which results in neglect of highly relevant frequencies of low amplitude.

Even though, both considered transformations show only minor deviations, the subsequent sections focus on PCA, as it yields more robust performance increments and offers a straightforward method of choosing well-functioning values of dimensionality reduction fraction based on the distribution of explained variances.

\subsection{Comparison of Training Data Noise and Dimensionality Reduction}
\label{chap:noise_dr_similarity}
We tested the ability of parallel reservoir computing to precisely predict time series of the KSE when trained only on noisy training data.
That is, the reservoirs are trained using input and output KSE time series that are perturbed with Gaussian noise of predefined signal to noise ratio
$ \mathrm{SNR}_{\mathrm{dB}} = 10 \log_{10}(P_{\mathrm{signal}}/P_{\mathrm{noise}}), $
where $P_{\mathrm{signal}}=1.71\pm0.01$ and $P_{\mathrm{noise}}$ denote the mean power of signal and Gaussian noise, respectively.
The signal power is calculated by averaging the power $u^2(x,t)$ of the KSE-variable over the whole spatial- ($[0,L]$) and temporal domain ([$0,T_{\mathrm{train}}]$) for all trainings datasets.

Subsequently, reservoirs are evaluated and best performing hyperparameters are selected using clean, i.e. noise-free, input data for the transient/ washout time and by comparing reservoir predictions to clean ground truth KSE time series.

\begin{figure}[t]
\includegraphics[scale=1.0]{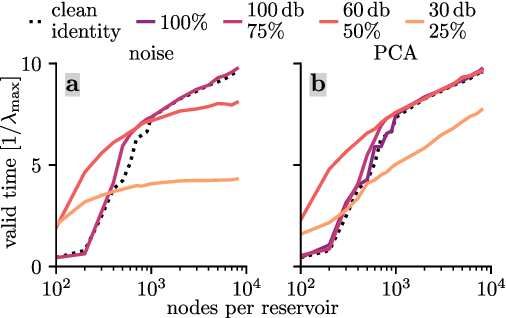}
\caption{
     \textbf{Weak training data noise and dimensionality reduction have similar effects on prediction performance for small reservoirs.}
     In Panel~\textbf{a} predictions with noisy data (\textit{coloured curves}) are compared to predictions with clean training data (\textit{dotted curves}), showing performance gains for small reservoirs and deterioration for large reservoirs.
     Panel \textbf{b} shows similar performance increments for small reservoirs and less severe decreases in performance for large reservoirs, comparing dimensionality reduced parallel reservoirs with PCA modes (\textit{coloured curves}) to the untransformed baseline (\textit{dotted curves}).
     All results use $\parres=32$ parallel reservoirs.
    }
\label{fig:noise_drrc_comparison}
\end{figure}

Figure~\ref{fig:noise_drrc_comparison} compares the influence of training data noise on parallel reservoir computing (\fig\ref{fig:noise_drrc_comparison} \textbf{a}) to dimensionality reduced parallel reservoir computing using a PCA (\fig\ref{fig:noise_drrc_comparison} \textbf{b}).
Here, noise levels are chosen based on the close agreement between the distribution of Gaussian noise and the deviation distribution of clean data to PCA-reduced data, i.e. $\lintrafo^{-1}\prodr\order\lintrafo$ (see appendix \fig~\ref{fig:noise_distribution}).
Qualitatively similar improvements in prediction performance can be observed for small node numbers when comparing noise of $\mathrm{SNR}_{\mathrm{dB}}=30$ to dimensionality reduction of $\dimred=25\%$, $\mathrm{SNR}_{\mathrm{dB}}=60$ to $\dimred=50\%$, and  $\mathrm{SNR}_{\mathrm{dB}}=100$ to $\dimred=75\%$.
We conjecture that similarities in performance between the two cases can be explained using the same reasoning:
In both cases, the linear readout $\wout$ is optimised using as input slightly perturbed dynamics of the KSE states and their dynamical impact on the driven reservoir states. 
Similar to previous discussions \cite{wikner_stabilizing_2024}, a stabilising response to deviations from the invariant set of the coupled system of KSE and reservoir is learned. 
Thereby, prediction performance is improved for small reservoirs, that often suffer empirically from instabilities in iterative applications.
Notably, this stabilisation of iterative predictions cannot be achieved by optimisation of the Tikhonov--regularisation parameter $\reg$, which agrees with findings of \cite{wikner_stabilizing_2024}.

In the limit of large reservoirs differences between the impact of noise and dimensionality reduction are apparent.
Here, the training data noise results in a reduced maximal valid time, i.e. an saturation to smaller limits of best performances for large reservoirs. 
Dimensionality reduction methods do not exhibit such saturation for up to $\nodes=8000$ nodes and across all tested dimensionality reduction fractions $\dimred$.
We attribute this difference in performance to the fact that data is perturbed differently in the two cases.
Training data noise perturbs both, KSE states used as reservoir input $\inp$ and as ground truth output states $\outp$, in all dimensions isotropically, whereas dimensionality reduction only perturbs input states along dimensions corresponding to truncated PCA modes, i.e. dimensions that explain little of the dynamics' variance.

\subsection{Counteracting Training Noise via Dimensionality Reduction}
\label{chap:noise_mitigation}

In this section, we analyse the impact of dimensionality reduction for parallel reservoir computing when applied to noisy time series.

For all tested noise levels, i.e. $\mathrm{SNR}_{\mathrm{db}}\in \{100, 80, 60, 40, 30, 20, 10\}$, sufficiently large dimensionality reduction, $\dimred \geq 50\%$, generally preserves prediction performance across all reservoir sizes compared to noisy predictions without dimensionality reduction. 
Exceptions to this finding are few isolated cases with minor performance decreases ($<5\%$), which lie well within error margins. 
For small reservoirs, dimensionality reduction consistently yields additional performance increments across all noise levels.
These gains are minor compared to the multi-fold performance increases observed for clean training data (compare \fig\ref{fig:LatentParRes_rel} \textbf{a} and \textbf{c}) and are smallest at an optimised noise level of $\mathrm{SNR}_{\mathrm{db}}=60$.
However, even under this condition, where substantial performance gains for small reservoirs arise from noise alone (compare \fig\ref{fig:noise_drrc_comparison} \textbf{a}), additional improvements of mean valid times of up to $30\%$ can be achieved by dimensionality reduction (e.g. $\tvalid\approx1.92 \to \approx2.50$ for $100$ nodes at $\dimred=50\%$).

\begin{figure}[t]
\includegraphics[scale=1.0]{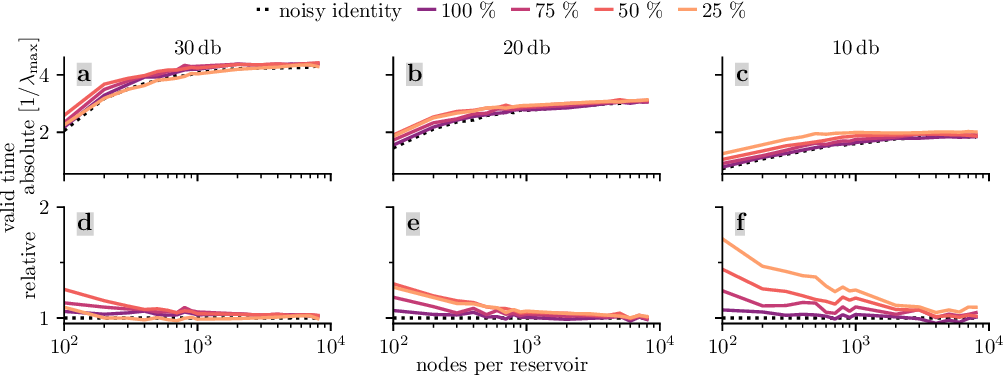}
\caption{\textbf{Dimension reduction methods mitigate the degrading effects on performance of strong training data noise.}
    Panels \textbf{a}, \textbf{b} and \textbf{c} show the deterioration of (absolute) prediction performance for increasing training noise levels of $30\,\mathrm{db}$, $20\,\mathrm{db}$ and $10\,\mathrm{db}$, respectively. Dotted curves denote prediction performance from noisy training data without dimensionality reduction, coloured curves from noisy training data with dimensionality reduction.
    Corresponding relative performance $\fper$ is shown in panels \textbf{d}-\textbf{f}.   
    Using dimensionality reduction yields general improvements in prediction performance ($\fper>1$), which are greatest for small reservoirs and strong training data noise
    All depicted results use $\parres=32$ parallel reservoirs.
    }
\label{fig:noise_robustness}
\end{figure}

For strong noise, which might occur in experimental measurement data,
\fig\ref{fig:noise_robustness} illustrates the deterioration of performance with increasing noise levels (see \fig\ref{fig:noise_robustness}~\textbf{a}-\textbf{c}). 
The corresponding relative performance $\fper$ (see \fig\ref{fig:noise_robustness}~\textbf{d}-\textbf{f}), i.e. comparing performances with noisy training data and dimensionality reduction (\textit{coloured curves}) to performances with noisy data and no dimensionality reduction (\textit{dotted curves}), highlights general improvements by using dimensionality reduction, i.e. $\fper>1$. 

Across all noise levels, in particular small reservoirs benefit from additional dimensionality reduction,

where relative performance improvements increase with stronger noise levels. 
This suggests that in the limit of strong training data noise, dimensionality reduction has a further, noise-mitigating role, by effectively filtering the noisy data.

\section{Discussion and Conclusion}

\label{chap:conclusion}

Within this paper, we quantitatively confirm that the established approach of parallel reservoir computing \cite{lu_reservoir_2017, pathak_using_2017, pathak_model-free_2018, zimmermann_observing_2018, herzog_reconstructing_2021, baur_predicting_2021, goldmann_learn_2022} 
significantly improves prediction performance.
Thereby, the approach allows for smaller reservoirs while maintaining the same mean prediction performance.
We further demonstrate intrinsic limitations of this method, namely that performance improvements by adding more and more parallel reservoirs gradually diminish to negligible levels.

Therefore we consider latent space predictions as an alternative method to reduce the input dimensionality.
While the sole use of latent space predictions has limited success in enhancing prediction performance, we find that  the combined approach of dimensionality-reduced parallel latent space predictions, enables additional improvement of predictions using small reservoirs. 
Thus, the combined approach effectively works as a downsizing tool for predicting chaotic dynamics with even smaller reservoirs.

Systematically testing the combined approach of dimensionality-reduced parallel latent space predictions with noisy training data, we find that improved performance consistently occurs for small reservoirs across all noise levels and that dimensionality reduction mitigates the deteriorating effects of strong noise levels.

We attribute the enhanced prediction performance of the parallel reservoir computing approach to its abilities to effectively make use of spatially-decoupled local states \cite{parlitz_prediction_2000}, reducing complexity and dimensionality of the input of each individual reservoir. 
This ability is inherently limited, as each parallel reservoir requires a sufficiently large local neighbourhood to capture the relevant coupled dynamics.
We therefore postulate, that the combined approach is successful because it reduces the input dimensionality of each reservoir to a minimum by, firstly, enforcing decoupled reservoir dynamics, which makes use of decoupled spatio-temporal dynamics and, secondly, utilising low-dimensional latent space representations of the local state, effectively removing local redundancies.
Importantly, the latter dimensionality reduction of the local state, only slightly perturbs the clean input signal in dimensions where little of the dynamics is explained, thereby stabilising iterative predictions similarly to the effect of training noise \cite{wikner_stabilizing_2024}.

These mechanisms, however, also highlight specific prerequisites and limitations of the approach:
Successful dimensionality reduction requires a system whose attractor dimension is considerably smaller than the system dimension, and a latent space transformation that preserves the local structure of the attractor, while allowing only minor perturbations of the signal.
In the KSE case, high linear autocorrelation between neighbouring pixels allows both tested linear dimensionality reduction methods to meet these requirements and thereby perform similarly well with only minor differences.

Nonetheless, the combined approach of parallel latent space predictions introduces three additional hyperparameters -- the number of parallel reservoirs, the size of the neighbourhood, and the dimensionality reduction fraction -- as well as the choice of the transformation.
While the introduction of new hyperparameters must be considered carefully, as it aggravates the often complex hyperparameter optimization task, our results suggest simple rules to guide their selection.
Namely, increasing the number of parallel reservoirs does not decrease prediction performance, leaving the user with an straight forward choice of taking as many parallel reservoirs as computationally achievable.
The neighbourhood size should be chosen as small as possible while ensuring that uncorrelated information from the surrounding is included.
In practice, this means it is preferable to select a slightly larger neighbourhood rather than risk missing relevant information.
Lastly, the PCA dimensionality reduction fraction can be chosen according to the distribution of explained variances, including all principal components with significant contribution to the cumulative explained variance.

For time series predictions of high-dimensional dynamical systems, classical reservoir computing often performs poorly.
This is partly because reservoirs, which are too small for their input system, struggle to extract relevant features from high-dimensional input data often leading to unstable iterative predictions.
Specifically, the limited size and random structure of input- and adjacency matrix, is badly suited to efficiently make use of globally decoupled and locally strongly-correlated input variables.
In addition, the use of Tikhonov-Regularisation has only limited success in stabilising iterative predictions \cite{wikner_stabilizing_2024}, which is in particular a problem for small reservoirs.
While these limitations can be overcome for the one--dimensional KSE by using larger reservoirs, this comes at the cost of increased runtime and memory.
Within this paper, we systematically analysed an alternative: By exploiting known properties of spatio-temporal systems --- namely, that no long range effects drive the dynamics and that the spatially-extended variable is smooth in space ---, we consistently achieve increased prediction performance of small reservoirs.

While downsizing is optional for 1D systems, it becomes essential for higher-dimensional systems, where required reservoir size and computational cost grow rapidly.
The challenge of high-dimensional reservoir input grows exponentially with the spatial dimensionality of the input domain, making effective dimensionality reduction increasingly important.
The presented approach, based on low-dimensional latent space representations of local states, is expected to scale well to higher-dimensional domains, as spatial decoupling and strong spatial correlations typically exist in all directions, opening possibilities for significant dimensionality reductions.

Going forward, the generality of improved performance and suggested hyperparameter rules need to be tested with other systems and especially higher-dimensional spatial domains. 
Thorough hyperparameter optimisation and performance evaluation for two- and three-dimensional spatio-temporal systems thus represent important next steps to fully assess the potential of dimensionality-reduced parallel latent space predictions.

\acknowledgements 
We thank Sebastian Herzog and Kai-Uwe Hollborn for scientific discourse during an early stage of the project.
GW acknowledges funding through a fellowship of the IMPRS for Physics of Biological and Complex Systems.
LF and UP thank Stefan Luther for supporting their research.
This work used the HPC system Raven at the Max Planck Computing and Data Facility and the Scientific Compute Cluster at GWDG, the joint data center of Max Planck Society for the Advancement of Science (MPG) and University of Göttingen.\\

\noindent
\textbf{Authors' Contribution}

LF and GW performed simulations and wrote the first draft of the manuscript. UP and GW conceptualized and supervised the project. LF, GW, and UP revised the manuscript.
\\

\noindent\textbf{Data availability statement}

Source code for this manuscript is available on GitHub \cite{DRRC-repo}.
Data used in this manuscript are available from the authors upon reasonable request.

\appendix
\renewcommand\thefigure{\thesection.\arabic{figure}}
\setcounter{figure}{0}
\section{Physics--Informed Weight Matrices}
\label{app:physics_informed_weightmatrices}
In the following, the structure of weight matrices of one reservoir, equivalent to $\parres$ parallel reservoirs, is illustrated for a one--dimensional spatio-temporal system where each parallel reservoir relies solely on predictions of adjacent reservoirs, i.e. $\coredim > 2\neidim$.
Therefore, let $\winpar, \wadjpar, \woutpar$ denote the input-, adjacency- and output matrices of the $\parind$-th parallel reservoir, respectively, where $\winpar \in \real^{\nodes\times\indim}$, $\wadjpar \in \real^{\nodes\times\nodes}$ and 
$\woutpar \in \real^{\coredim\times(1+\nodes+
\indim)}$ for all $\parind\in\{1,..,\parres\}$. 
Further, let $\winpar=(\winleft|\wincore|\winright)\in\real^{\nodes\times3\coredim}$ be the decomposition of input matrices in mappings of the left neighbourhood, the core and the right neighbourhood variables, for $\winleft$, $\wincore$ and $\winright$ respectively. 
Here, without loss of generality ($\coredim > 2\neidim$) we assume that $\winleft, \wincore, \winright \in \real^{\nodes\times\coredim}$, by adding sufficiently many columns of zeros to $\winleft$ and $\winright$.
The decoupled inner dynamics, i.e. non interacting reservoir states, can be enforced by choosing a block diagonal structure of the adjacency matrix $\wadj$.

Similarly the input matrices can be arranged into a block diagonal matrix with overlapping blocks.
Hence, equivalent reservoir dynamics of one large reservoir is given by
\begin{align}
\wadj &= 
\begin{pmatrix}
\wadjpar[1] & 0       & \cdots  & 0 \\
0       & \wadjpar[2] & \cdots  & 0 \\
\vdots  & \vdots  & \ddots  & \vdots \\
0       & 0       & \cdots  & \wadjpar[\parres]
\end{pmatrix}, \\
\win &= \begin{pmatrix}
\wincore[1] & \winright[1] & \cdots  & \winleft[1] \\
\winleft[2] & \wincore[2] &\winright[2]   & \cdots\\
\vdots  & \vdots       & \ddots & \vdots \\
\winright[\parres]  & \cdots       & \winleft[\parres] & \wincore[\parres]
\end{pmatrix}.\notag
\end{align}
Similarly, the use of parallel reservoirs enforces conditions on the linear superposition matrix $\wout$.
Namely, with an extended state vector $\esv = [\bin, \res,\inp]\trans$, the output matrix consists of a block diagonal structure for weights acting on the reservoir states, i.e. $(\wout)_{i,j}$ with $j\leq \nodes\parres$, and an overlapping block diagonal for weights acting on the input, i.e. $(\wout)_{i,j}$ with $j>\nodes\parres$.
The here presented construction is designed for one--dimensional systems, similar decompositions of matrices exist for arbitrary system dimensions $\spacedim\geq1$. 
The use of block diagonal reservoir structures is used in and analysed for the prediction of low dimensional systems of ODEs in \cite{ma_novel_2023}.

\section{Generality of Qualitative Results}
\label{sec:Generality_of_Qualitative_Results}
\begin{figure*}[ht]
\includegraphics[width=\textwidth]{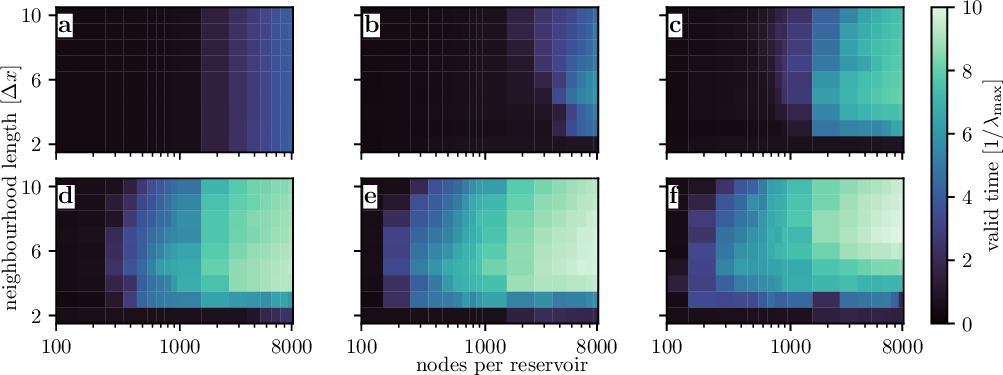}
    \caption{
        \textbf{Optimal neighbourhood length $\neilen$ depends on node number and number of parallel reservoirs.}
        The Figure summarizes the dependence of performance on neighbourhood size and node number portraying results for $\parres\in\{1,2,4,8,16,32\}$ parallel reservoirs in \textbf{a}-\textbf{f}, respectively. 
    }
    \label{fig:ParRes_nei_all}
\end{figure*}

The dependence of prediction performance on number of parallel reservoirs and neighbourhood size is shown in \fig\ref{fig:ParRes_nei_all}. 
For all reservoir sizes $N$ and numbers of parallel reservoirs $\parres>1$, one observes an optimal neighbourhood length $\neilen$. 
Specific values of this optimal neighbourhood length depend on node number and number of parallel reservoirs but are mostly in $[3\Delta x,8\Delta x]$.

\begin{figure*}[ht]
\includegraphics[width=\textwidth]{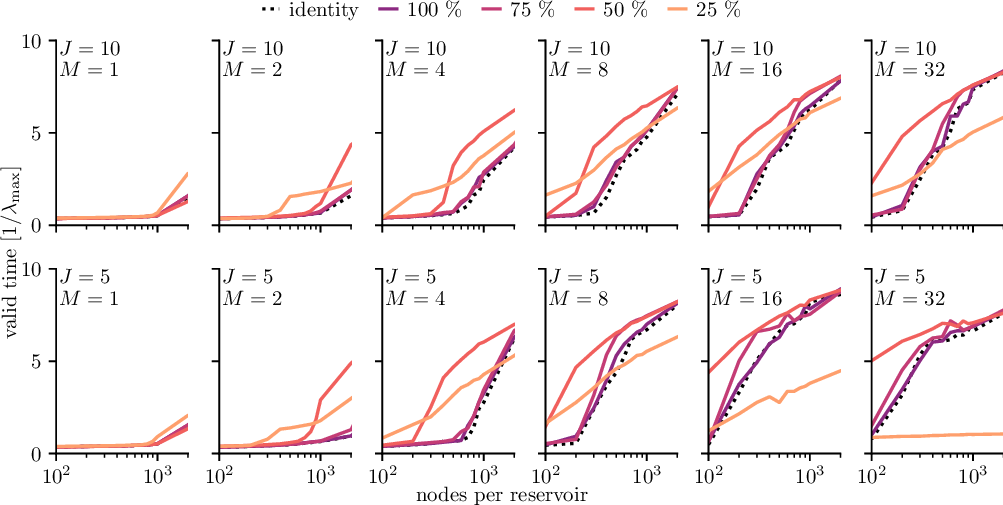}
    \caption{
        \textbf{The performance improvement of parallel dimensionality-reduced latent space predictions is independent of optimization of the number of parallel reservoirs and the  neighbourhood length $\neilen = \neiind \Delta x$.}
        The figure summarizes the dependence of performance on node number $\nodes$ (\textit{x-axes}), number of parallel reservoirs $\parres$ and neighbourhood length $\neilen$ and dimensionality-reduction fraction $\dimred$ (\textit{colour}) using the PCA as transformation. Note the great performance for small reservoirs in the case of dimensionality-reduced parallel latent space predictions with $\neilen=5\Delta x$.
    }
    \label{fig:drrc_nei_compare}
\end{figure*}

Figure \ref{fig:drrc_nei_compare} shows that the discussed improvement of performance by dimensionality-reduced parallel latent space predictions is not constrained to specific numbers of parallel reservoirs $\parres>1$ and neighbourhood length $\neilen=\neiind\Delta x$.
Further, the figure shows that by decreasing the neighbourhood length to $\neilen=5\Delta x$ even greater performance improvements are observed for small reservoirs.

\section{Training Data Noise}
\label{app:noise}
\begin{figure}[h!]
\includegraphics[scale=1.0]{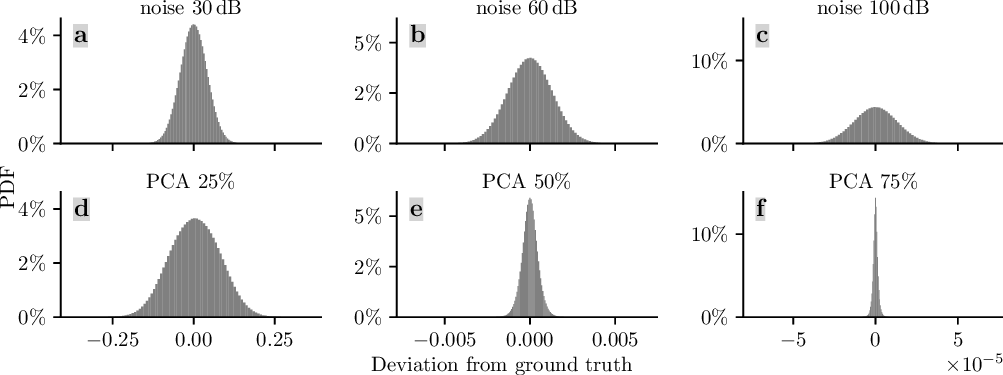}
\caption{
    \textbf{Dimensionality Reduction is a loss of information similar  to adding Gaussian noise.}
    The figure shows the distribution of Gaussian noise (\textit{top}) with different noise levels ($30\,$db, $60\,$db and $100\,$db) and the 
    deviation distribution of PCA-reduced data in real-space, i.e. $\lintrafo^{-1}\prodr\order\lintrafo$, to clean data. 
    to ground truth signal from dimensionality reduced PCA modes back transformed to real-space (\textit{bottom}). Similarities in magnitudes between the two rows are apparent.}
    \label{fig:noise_distribution}
\end{figure}

In the two sections evaluating the influence of training data noise, we applied different heuristics to chose considered noise levels. 
To test robustness (see \sect~\ref{chap:noise_mitigation}), we applied noise levels of $30$, $20$, and $10\,$dB, representative of strong noise that might occur in experimental data.
To compare prediction performance between dimensionality reduction and training data noise, we used noise levels of $30$, $60$, and $100\,$dB. 
These levels were chosen because of a close match of scales between the Gaussian noise distributions and the deviation distribution observed between noise free and PCA-reduced data (applying $\lintrafo^{-1}\prodr\order\lintrafo$).
Figure \ref{fig:noise_distribution} shows the distributions side by side: top row (panels \textbf{a–c}) shows noise levels $30$, $60$, $100\,$dB; bottom row shows corresponding PCA reductions of $25\%$, $50\%$, and $75\%$.
Of particular interest are comparisons between panels \textbf{a,b} and \textbf{c,d}, respectively, where performance changes are most pronounced.
Although differences are visible within each column, the overall alignment becomes apparent when comparing amplitudes and the scale of deviations across panels in each row.
Notably, the $\dimred=75\%$ reduction (Panel \textbf{d}), where only minor differences in performances are observed, shows the largest deviation, in shape and magnitude.

\section{Numerics}
\label{chap:numerics}
\subsection{Solving the KSE.}
Equation~\eqref{eq:ks} is best solved using a spectral method, such that it can be rewritten as
\begin{equation}
  \partial_t \mathcal{F}\{u\}(k, t) = \frac{\mathrm{i} k }{2} \mathcal{F}\left\{ \mathcal{F}^{-1} \left\{ \mathcal{F}\{u\}\right\}^2\right\}(k,t) + (k^2 - k^4) \mathcal{F}\{u\}(k,t) \,. 
\end{equation}
Here $\mathcal{F}\{u\}(k,t)$ denotes the Fourier transform of the field $u(x,t)$.
Note that this PDE is the sum of a non-linear and a linear operation on $u$, such that both can be discretised in time separately.
In this work we use a Crank-Nicholson and an Adams-Bashforth scheme for the linear and non-linear parts \cite{Press-Numericalrecipes}, respectively.
Parameters can be found in \tab\ref{tab:parameter} and \ref{tab:numerical_params}.

\begin{table}[h]
  \caption{
    Other parameters used. 
    The Lyapunov time $ \lambda_\mathrm{max}$ was calculated with code supplied with \cite{datseris_nonlinear_2022}.
  }
  \label{tab:numerical_params}
  \begin{tabular}{lr}
  \toprule
  $ \lambda_\mathrm{max} $ & 0.095\\ 
  $\min AC$ & $8.3 \Delta x$ \\ 
  $AC_0$ & $4.6 \Delta x$\\
  $e$ & 0.5 \\
  $L$ & 60 \\
  $D$ & 128 \\
  $m_\mathrm{train}$ & 50000 \\
  $m_\mathrm{trans}$ & 100 \\
  $\#_\mathrm{initialisations}$ & 10 \\ 
  $\#_\mathrm{evaluations}$ & 50 \\ 
  \bottomrule
  \end{tabular}
\end{table}

\subsection{Implementing parallel latent space predictions.}
To ensure a simple generalisation of the implementation to parallel latent space predictions (see \sect\ref{chap:ext_lat_intro}), in numerical implementations we also train the predictions of neighbourhood cells. 
The prediction of neighbourhood cells are assumed to be flawed and neglected in iterative predictions.
Note, that this does not effect the training of predictions of core cells, as individual rows of $\wout$ are optimised independently.
\clearpage
\bibliography{references}

\end{document}